\documentclass[11pt, letterpaper]{article}
\usepackage{amsmath, graphicx,hyperref}
\usepackage{amsthm,latexsym, amsfonts,amssymb,amstext}
\usepackage{fullpage} 
\usepackage[usenames]{color}
\usepackage{csquotes}
\usepackage{multicol}
\usepackage{microtype}
\usepackage[framemethod=tikz]{mdframed}
\usepackage{xcolor}
\usepackage{changes}
\usepackage[margin=20pt,
font+=small,labelformat=parens,labelsep=space,
skip=6pt,list=false,hypcap=false
]{subcaption}
\usepackage{wrapfig}

\title{Elastic Cash}

\author{Anup Rao \\University of Washington\\{\tt anuprao@cs.washington.edu}}

\begin{document}
	\maketitle
\begin{abstract}
Elastic Cash is a new decentralized mechanism for regulating the money supply.  The mechanism operates by modifying the supply so that an interest rate determined by a public market is kept approximately fixed. It can be incorporated into the conventional monetary system to improve the elasticity of the US Dollar, and it can be used to design new elastic cryptocurrencies that remain decentralized.
\end{abstract}
\section{Introduction}
Money is as old as recorded history, and yet it continues to evolve. Even the mighty US Dollar has been repeatedly updated over the last 200 years. The recent emergence of Bitcoin and other cryptocurrencies is another step in that evolution, and it prompts us to revisit the mechanisms that ensure the desirable properties of money. An important property of money is its \emph{elasticity}:  money is elastic if the money supply increases in response to  demand.  In this work, I present a new decentralized  mechanism to  ensure the elasticity of money.

The US Dollar is elastic, but it uses a convoluted system to achieve its elasticity. The Dollar system was not designed from first principles; it was iteratively amended in response to financial crises. Elasticity is currently  achieved by the combined actions of many institutions: banks and non-banks, public and private, domestic and international. No single entity is entirely in charge of the money supply, and a relatively small number of investor-owned institutions have undue influence. This system computes the demand for money without much  transparency, and the money that is created is not distributed fairly. I discuss the mechanics of the US Dollar system and its limitations  in Section \ref{dollar}.

In contrast, Bitcoin was designed to be inelastic. Bitcoin caps the total possible supply at $21$M, and the available supply,  currently about $19.2$M, will slowly increase until it hits this limit. Satoshi Nakamoto, the creator of Bitcoin, criticized  conventional mechanisms for achieving elasticity in an early forum post:
\begin{displayquote}
The root problem with conventional currency is all the trust that's required to make it work. The central bank must be trusted not to debase the currency, but the history of fiat currencies is full of breaches of that trust. Banks must be trusted to hold our money and transfer it electronically, but they lend it out in waves of credit bubbles with barely a fraction in reserve.
\end{displayquote} Nakamoto eliminated the need for the kind of infrastructure used to ensure Dollar elasticity by simply choosing to make Bitcoin inelastic. Additional technological innovations in Bitcoin further eliminated the need for any trusted central authority to carry out transactions. Bitcoin inspired many other cryptocurrencies, incuding stablecoins that are pegged to the US Dollar. Some of these currencies are elastic, but their elasticity is ultimately based on the elasticity of the US Dollar, so they inherit the limitations of US Dollar elasticity. I discuss elasticity in the context of cryptocurrencies in greater detail in  Section \ref{crypto}.

To illustrate the negative consequences of inelasticity, consider the trajectory of home prices denominated in Bitcoin over a long period. As the population grows, the demand for houses is sure to grow. Even if the supply of houses keeps  pace, home prices must fall, because the supply of Bitcoin cannot keep pace. Certainly, if the number of concurrent transactions involving home sales grows and the prices of houses remain stable, the amount of Bitcoin involved in these transactions would have to grow, yet it cannot grow beyond $21$M. So, a fixed money supply leads to falling prices in a growing economy, even if the underlying supply and demand of items keep pace with each other.

At the other extreme, if the money supply is increased excessively, the currency is debased; the excess supply leads to rising prices and inflation. So, a  mechanism for correctly setting the money supply is essential, because this is the foundation upon which stable prices are built. As much as possible, prices should reflect the tension between the supply and demand for items, and nothing else. But how much money is too much? It is not just a matter of trust, the problem is that the appropriate supply is difficult for anyone to calculate! Is there a principled method to compute the supply, and a fair way to create new money? This work gives my answers to these important questions.

Ideally, a mechanism for achieving elasticity should be transparent. It should not rely on the judgment or integrity of small groups of people, or a few  institutions. I will describe a new decentralized  mechanism called \emph{Elastic Cash} that enjoys these features. Elastic Cash solves two problems at the same time:
\begin{itemize}
	\item It gives a fully decentralized mechanism to compute whether the demand for money has increased or decreased.
	\item It gives a fair mechanism to increase and decrease the supply of money.
\end{itemize}

Here I will give a high-level description of Elastic Cash; a full description is in Section \ref{new}. I will refer to the central bank and the algorithm  as \emph{cash authorities}. At the heart of the mechanism is a new financial contract issued by the cash authority called a \emph{cashbond}, and a public market that allows anyone to buy and sell cashbonds using public auctions in the market. Each cashbond can be redeemed for \$1 on a specific date, so it executes a risk-free loan from the holder of the cashbond to the cash authority. The loans are risk-free because the cash authority can always create new money to repay the loans. Perhaps it is counterintuitive, but the purpose of the market in cashbonds is \emph{not} to give the cash authority a way to borrow money; instead, the function of the market is to use the trading activity of participants to compute the   \emph{risk-free rate of interest}, which can be computed from the prices of cashbonds in the market. The mechanism requires that cashbonds can only be transfered by selling and buying them in the market. For example, no entity should be  permitted to use cashbonds as collateral to borrow money. This forces participants to liquidate cashbonds when they need money, and keeps the mechanism informed about the demands for money.

It is market participants who determine the money supply in Elastic Cash.  The supply is increased or decreased according to transparent rules ensuring that the risk-free rate of interest  remains approximately fixed. Trade in cashbonds leads to fluctuations in the rate of interest, and the mechanism responds by creating corresponding fluctuations in the money supply. Intuitively, the risk-free rate of interest encodes the cost of renting money. By regulating the supply of money to keep the cost fixed, the mechanism ensures that the supply stays in equilibrium with the demand for liquidity. When the supply is increased, market participants acquire any newly-generated money. The supply is decreased by incentivizing  market  participants  to exchange their money for cashbonds.  Anyone can participate in this market, and money is distributed to or taken from participants according to transparent rules, so no single entity can control the flow of money. 

Elastic Cash can be implemented within the framework of conventional currencies like the US Dollar by creating regulations that require the central bank to implement the market for cashbonds and generate money according to the rules of the mechanism. It can be implemented in the framework of cryptocurrencies by setting up a distributed algorithm to implement the market for cashbonds using the blockchain. Money is generated by the algorithm according to the rules of the mechanism. So, one can obtain the positive features of traditional currencies and  cryptocurrencies in addition to the elasticity of Elastic Cash.

\paragraph{Outline of this paper}
In Section \ref{dollar}, I give more details about how the US Dollar achieves its current elasticity, before turning to explain how known elastic cryptocurrencies inherit the limitations of the Dollar in Section \ref{crypto}. I describe the new mechanism in Section \ref{new}. I discuss how the new mechanism can be incorporated into the Dollar system in Section \ref{newdollar} and how it can be implemented on the blockchain to give new  cryptocurrencies in Section \ref{newcrypto}.

\section{US Dollar elasticity: the Fed, banks, and shadow banks} \label{dollar}

\begin{figure}[t]
	\centering
	\includegraphics[width = 17cm]{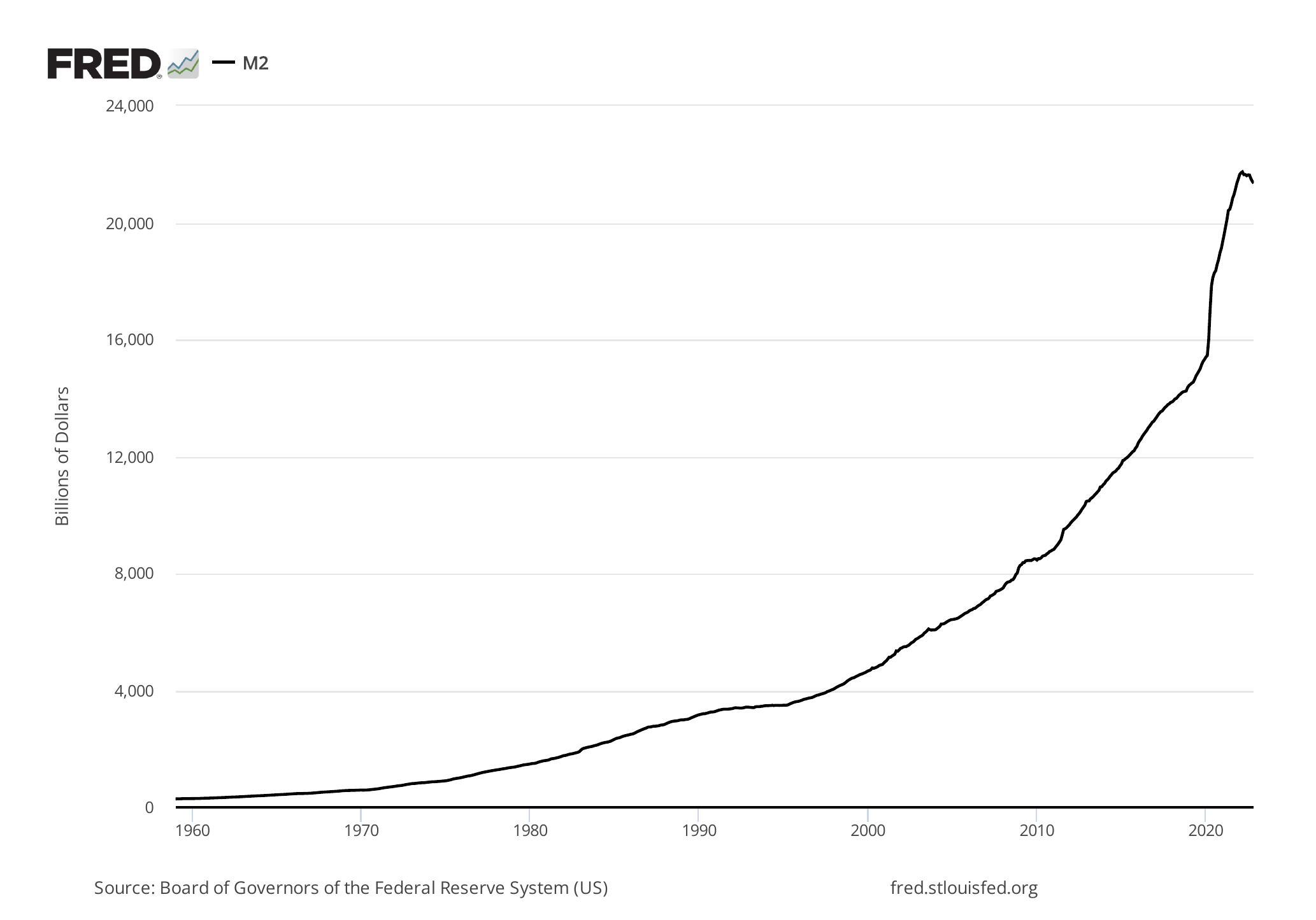}
	\caption{M2, a measure of the  supply of US Dollars} \label{m2}
\end{figure}

Figure \ref{m2} shows how the supply of US Dollars held as deposits has changed over time. In this section I briefly review the history and mechanics of the US Dollar system. I  recommend the following excellent sources for additional background on the history of the Dollar system  \cite{fedweb, fedunbound}, and the book \cite{makingmoney} for a longer history of the technology of money.

The US Dollar is the fiat currency with the most advanced infrastructure for maintaining elasticity. The supply of US Dollars has  consistently been deemed too important to be left entirely in the control of a government agency. Instead, we have tried to decentralize the creation of money by developing a system where money is created by investor-owned private entities. These include banks and so-called shadow banks --- non-bank financial institutions that are able to create Dollars. This privately created money is then backstopped by the central bank, which is the Federal Reserve, or Fed in the US. The Fed is governed by laws written by Congress, but these laws are somewhat ambiguous about the Fed's powers, and Congress has repeatedly made adjustments. 

There are three major ways in which the Fed can influence the money supply:
\begin{enumerate}
	\item It can change the Fed funds rate, which influences all borrowing in the financial system, and changes the incentives for private entities to create money.
	\item It can elevate a new kind of asset to the stature of the US Dollar, which gives a new class of institution the power to create Dollars.
	\item It can directly buy or sell assets under a Quantitative Easing/Tightening  program.
\end{enumerate}

Today, there are at least four kinds of private institutions that are able to generate US Dollars. It is helpful to view the evolution of the US Dollar system according to the events that elevated the instruments created by these institutions to the stature of US Dollars:

\begin{description}
\item [Commercial bank deposits]
By the early 1900s, deposits of US Dollars were being issued by a number of investor-owned private banks.  These deposits were treated by depositors as equivalent to US Dollars, even though the banks were issuing loans by creating deposits that did not correspond to cash reserves. This led to a  run on the  banks in 1907, and the Fed was created in 1913 to solve the problem. The Fed was given the power to backstop the money created by commercial banks by  lending to the banks. This meant that deposits could always be exchanged for money created by the Fed. This solved the problem of bank runs, but also elevated bank deposits to the stature of cash issued by the Fed, and effectively meant that commercial banks were given the authority to create legitimate US Dollars in the process of issuing loans. Currently, about $\$17$T is held in commercial bank deposits.

\item[Reverse purchase agreements (repos)] In the 1950s, broker dealers began to enter the banking industry using a financial instrument called  reverse purchase agreements, or repos. Repos allow  dealers to borrow money from cash providers using government securities as collateral. Cash providers began to treat the repos they had purchased from dealers as equivalent to cash. By itself, such an arrangement does not create any new money, because if repos crash in value, then they cannot actually be exchanged for cash. However, the properties of repos were substantially altered when the Fed decided that the repo market was important enough to backstop dealers by giving them access to overnight loans. This meant that the private holders of repos were guaranteed that repos could always be exchanged for US Dollars via the Fed's repo facility, and so repos were elevated to the same stature as cash and bank deposits. In 1991, Congress reduced restrictions on the Fed to make it even easier to backstop the repo market. Effectively, broker dealers were given the power to create new US Dollars. Currently, the size of the repo market is about $\$4$T.

\item [Eurodollars] Foreign companies including both banks and non-banks (e.g. insurance companies) have been issuing financial instruments called eurodollars that can be redeemed for US Dollars. These eurodollars were not initially backed by actual Dollars. The oil shock of 1973-1974 led to a run on eurodollars  that eventually brought down Franklin National Bank, a large domestic US bank that had access to the Fed's lending facilities. The Fed responded by promising to backstop eurodollars by providing actual US Dollars in the form of loans to  foreign central banks. This announcement ended the run on eurodollars, but  effectively permitted the banking systems of other countries to create deposits that could be exchanged for US Dollars. The eurodollar market has grown substantially since then. The size of the eurodollar market was estimated at about $\$13$T in 2016 \cite{eurosize}.

\item[Money market mutual funds] These funds emerged in the 1970s as investments whose share price was pegged at \$1. In reality, these funds held assets whose value could drop below the peg, and so it was not possible to guarantee the peg during a financial crisis. In response to the great financial crisis of 2008, the Fed began to backstop these funds using its Money Market Mutual Fund Liquidity Facility, and so elevated deposits in these  funds to the stature of US Dollars. Total assets in these funds is about $\$4.8$T.
\end{description}

In addition to recognizing new forms of the US Dollar, the Fed has resorted to buying assets in order to inject liquidity  under the \textbf{Quantitative Easing} program. During the Great Financial Crisis of 2008, the Fed kicked off this program by buying mortgage-backed securities and treasuries, which are loans to the US Treasury. Throughout the last decade, the Fed has continued to expand its balance sheet, mostly with treasuries. In 2020 the Fed once again bought significant quantities of these assets. Currently the Fed holds  about $\$8.5$T on its balance sheet.

\subsection{Limitations}
The history of the US Dollar is full of ad hoc amendments to maintain stability in the face of financial crises. Here are some limitations of the current system:
\begin{enumerate}
	\item At its heart, the problem is that there is currently no principled way to regulate the supply of money. This becomes apparent during times of financial crises, but the imbalance in supply is probably always  brewing, even in normal times. 
\item Although the Fed is nominally not in charge of the money supply, if the Fed funds rate has been lowered to $0\%$ and the Fed decides that more money is needed, the Fed has repeatedly chosen to inject money by buying  assets and sometimes propping up new kinds of financial assets. 
This goes against the principle of having a distributed system of money creation, and interferes in the functioning of financial markets. 
\item 
The Fed does not have the tools to distribute the money that it creates fairly. For example, money that is created via the Quantitative Easing program largely enters the system in the form of government spending and subsidized mortgages, and the decision to backstop repo protected broker dealers. There is no reason why newly created money should be tied directly to government spending, the mortgage industry or broker-dealers; this is an artifact of bad design. 
\item It makes sense to decentralize the power to create money, but the current system does this in a half-hearted way: the power remains concentrated in the hands of big players in the financial industry, who often get first access to the newly-created money. 

\item Because private financial institutions with the power to create Dollars know that the Fed will protect their financial instruments from the most negative consequences of their choices, their incentives are not aligned with creating the appropriate amount of money.
\end{enumerate}
Elastic Cash is meant to provide a clean, transparent, and principled mechanism to achieve robust elasticity by giving a truly decentralized way to compute the appropriate supply as well as a fair way to distribute new money. We do not need to cede control of the money supply to private institutions or foreign banks. The mechanism will generate new US Dollars when required, and financial institutions can obtain liquidity by participating in the mechanism, just like everyone else. Even if we want to leave the powers of the Fed to create new money undiminished, the scheme of Elastic Cash, at the very least,  gives a fair way to distribute new money. It  removes the ability of the finance sector to control the flow of the new money, and puts everyone on the same playing field.

Extricating ourselves from the current system and its vested interests is likely to be challenging, to say the least. Nevertheless, I describe a path to incorporating the new mechanism in Section \ref{newdollar}.

\section{Elasticity in Cryptocurrencies} \label{crypto}

Stablecoins are cryptocurrencies that are pegged to the Dollar by holding actual Dollars. These include Tether, USD Coin, and True USD. Every coin is backed by a US Dollar and  new coins can be created backed by new US Dollars. So, the elasticity of these coins is ultimately tied to the elasticity of the US Dollar, and an excessive increase in the supply of Dollars can lead to corresponding increases in the supply of these coins. These coins inherit the drawbacks of the US Dollar discussed in Section \ref{dollar}.

The Maker protocol \cite{maker} is a slightly different approach, where the currency is \emph{soft-pegged} to the US Dollar. Users are able to vote on what constitutes valid collateral, and currency is created against this collateral. Of course, the US Dollar or other currencies are involved in the peg, and so once again the supply of this money is tied to the supply of Dollars and other fiat currencies.

Similarly, the Frax system \cite{frax} provides cryptocurrencies with \emph{fractional} collateral, where the currency is not backed Dollars $1:1$, but still attempts to maintain a peg to the US Dollar. Once again, the supply of FRAX is ultimately tied to the supply of US Dollars. Another approach in the Frax ecosystem is the Frax Price Index (FPI) stablecoin \cite{fpi}. FPI attempts to maintain a peg to the consumer price index, which tracks the prices of a basket of goods denominated in US Dollars. Short of actually holding the constituent goods, there is no principled way to maintain this peg, and so it is possible that the coin will deviate from the peg during times of price instability.

As far as I am aware, there is no  cryptocurrency that is currently equipped with a native mechanism like Elastic Cash to maintain elasticity. I believe one can implement Elastic Cash in a cryptocurrency and create the first cryptocurrency with a self-contained decentralized mechanism for elasticity!  

\section{Elastic Cash: the details} \label{new}

Elastic Cash uses trade in cashbonds to determine a risk-free rate of interest. The money supply is regulated to ensure that this interest rate remains approximately fixed. Cashbonds are issued by the central bank (in the case of conventional currencies) or by the distributed algorithm (in the case of cryptocurrencies). I refer to these entities as \emph{cash authorities}. 

The contract  $\mathsf{cashbond}(d)$  promises that the cash authority will pay the holder of the contract \$1 on the date $d$. Let us reserve $d_0$ to denote the current date. On date $d_0$, the cash authority pays each holder of $\mathsf{cashbond}(d_0)$   \$1, and these contracts expire. 
Elastic Cash requires that the cash authority implement a public market in cashbonds.
On the date $d_0$,  contracts of the type $\mathsf{cashbond}(d)$ for $d > d_0$ will be available in the market for cashbonds maintained by the cash authority.

Cashbonds are a special class of asset, and they should not be treated like other securities. Elastic Cash requires that cashbonds can be generated and traded \emph{only} in the public market that is administered by the cash authority. Cashbonds are not transferable, meaning they cannot be exchanged outside of the public market, and they cannot be used as collateral for loans. Because of these restrictions, cashbonds cannot themselves play the same role as money. The purpose of these rules is to ensure that holders of cashbonds who desire liquidity will sell their cashbonds in the market and so keep the mechanism informed about the demand for liquidity. For the same reasons, trade in cashbonds should not be taxed. The transactions of buying and selling cashbonds should be viewed as similar to transactions that move money between savings accounts paying varying rates of interest, and treated similarly under the law.

\subsection{Risk-free rate of interest}
The price at which cashbonds trade implies interest rates for risk-free loans of varying durations. Let $\mathsf{rate}(t)$ denote the interest rate for duration $t$, where $t$ is measured in units of years. Note that $t$ can take fractional values. Let us write $\mathsf{price}(d)$ to denote the price at which  $\mathsf{cashbond}(d)$ last traded in the market. Then, if the current date is $d_0$ (again measured in years),  the prices of cashbonds can be used to compute implied interest rates according to the formula:
$$ \mathsf{price}(t+d_0) \cdot (1+\mathsf{rate}(t))^{t}= 1,$$ which implies that the interest rate can be expressed as
$$ \mathsf{rate}(t) = \mathsf{price}(t+d_0)^{-\frac{1}{t}}-1.$$
Because the loans executed by cashbonds are risk-free, the values  $\mathsf{rate}(t)$ capture something about the market's belief about the opportunity cost of making risk-free loans for duration $t$. 
Generally, one would expect $\mathsf{rate}(t)$ to be a monotone function of $t$, meaning that 
$\mathsf{rate}(t)>\mathsf{rate}(t')$ if $t>t'$, because loans of longer duration usually command higher interest rates. Moreover, if $t$ is much larger than $t'$, then we might expect $\mathsf{rate}(t)$ to have higher variance than $\mathsf{rate}(t')$, because predictions about the distant future can diverge much more than predictions about the immediate future. It is tempting to make inferences about the values of $\mathsf{rate}(t)$ based on the current treasury market, but I believe these markets are significantly different from each other, and so historic data about treasury yields is unlikely to capture how the risk-free rate of interest would behave in the cashbond market. I discuss the differences in Section \ref{discussion}.

These rates encode important information about the demand for liquidity. The goal of the mechanism is to regulate the money supply so that one of these rates is held approximately fixed. It makes the most sense to pick a rate for a relatively short duration, because these rates are likely to have the least variance. With that in mind, let $\tau$ denote a short time period, say $1$ week, so in years we have $\tau = 7/365$. The goal of the mechanism will be to keep 
$$\mathsf{rate}(\tau) \approx 0.02.$$
There is nothing special about $0.02$, except that it is convention for central banks around the world to use $2\%$ as the target rate of long term inflation.

Let us set $$p_- = (1+0.021)^{-\tau},$$ and $$p_+ = (1+0.019)^{-\tau}.$$
The goal of the mechanism will be to regulate the money supply so that $$p_- \leq \mathsf{price}(\tau+d_0) \leq p_+,$$
where again $d_0$ is the current date. This corresponds to keeping
$$ 0.019 \leq \mathsf{rate}(\tau) \leq 0.021.$$

\subsection{Using the market to regulate the money supply}
Participants in the cashbond market can put in orders to sell a specific number of cashbonds that they hold at a specific price, and can also put in orders to buy a specific number of  $\mathsf{cashbond}(d)$ at a specific price. The cash authority acts as a \emph{market maker} to match buy orders to sell orders and so conduct transactions at a specific price between market participants. Ideally, the market for cashbonds will support auctions\footnote{I will not commit to a specific style of auction here, though any implementation must carefully specify how the cash authority behaves as a market maker and what the rules of the auctions are.} for sellers to sell their cashbonds when needed. 

The cash authority will itself participate in this public market by buying and selling cash bonds in prescribed ways. The goal of the mechanism is to maintain $\mathsf{rate}(\tau)$ approximately fixed, and to keep the market in cashbonds sufficiently liquid, so that the money supply can be quickly adjusted based on changes to $\mathsf{rate}(\tau)$. 

It makes sense to pick a particular target distribution on outstanding cashbonds that is maintained during normal times.  If the current date is $d_0$, we say that $\mathsf{cashbond}(\delta+ d_0)$ has duration $\delta$. For example, the cash authority might aim to maintain the invariant that at any point in time, $1/4$ of the outstanding cashbonds have duration between $0$ and $1$ month,  $1/4$ have duration between $1$ month and  $1$ year, $1/4$ have duration between  $1$ year and $4$ years, and $1/4$ have duration between $4$ years and $10$ years. I discuss some considerations for how this distribution should be picked in Section \ref{discussion}.

Here is the proposed scheme for buying and selling cashbonds:
\begin{enumerate}
	\item The cash authority will buy and sell cashbonds to keep $\mathsf{rate}(\tau) \approx 0.02$. The cash authority will place a standing order to buy an infinite number of contracts $\mathsf{cashbond}(\tau + d_0)$ at price $p_-$,
	and a separate standing order to sell an infinite number of $\mathsf{cashbond}(\tau + d_0)$ contracts at price $p_+$.
	
	Because the cash authority is able to generate arbitrary amounts of both money and  cashbonds, it will always be able to satisfy any of the resulting transactions. This ensures that 
	$$p_- \leq \mathsf{price}(\tau+d_0) \leq p_+,$$
	as discussed above.

\item When cashbonds are redeemed for money, the cash authority will need to sell new cashbonds to restore the balance between money and cashbonds. The redeemed cashbonds should be replaced by selling new cashbonds at auction, picking the dates of the new cashbonds so that the overall distribution on duration is restored to the target distribution on durations, as much as possible.

	\item When the demand for money is high, we are likely to  reach the point where all of the available bonds  $\mathsf{cashbond}(\tau + d_0)$ have been purchased by the cash authority. In such times, the mechanism has run out of the means to inject money into the financial system at a fast enough pace according to rule 1.  This can be resolved by selling large quantities of  $\mathsf{cashbond}(2\tau + d_0)$ contracts at auction in the market. Market participants will be incentivized to buy these cashbonds and then sell them back after time $\tau$; at that time the cash authority itself will be willing to buy the cashbonds at price $p_-$. The net effect will be to inject money into the system, while preserving the number of outstanding cashbonds.
	
	The number of cashbonds sold in this process is a design choice. The goal  is to inject significant liquidity, so I would favor an exponentially escalating volume of sales. For example, the cash authority might first sell a quantity that corresponds to  $1\%$ of all outstanding cashbonds, and a week later escalate it to $2\%$, then $4\%$, and so on until the cashbond market returns to the state where  market participants are no longer willing to sell back the cashbonds of duration $\tau$ to the cash authority at $p_-$. These actions may temporarily distort the distribution on the  durations of outstanding cashbonds, but the distribution will be quickly restored when the new cashbonds are redeemed and rule 2 is applied.
	\end{enumerate}

These rules allow the money supply to rapidly adjust to the demand for liquidity. An actual implementation of Elastic Cash would need to resolve many smaller technical details. Let me now make a few comments and observations about the Elastic Cash mechanism as I have defined it.

\subsection{Discussion}\label{discussion}
Here I want to make a few observations about Elastic Cash as I have defined it:
\begin{description}
\item[Elastic Cash provides is a truly new mechanism] The Fed funds rate is a rate set by the Fed, while Elastic cash allows the risk-free rate of interest to float freely in a narrow range, and injects money only when needed to change the rate. This is a significantly different process. Elastic Cash is quite different from a system where the central bank simply allows deposits for all with interest rate $2\%$---such a scheme does not give the central bank a method to inject large amounts of money when the liquidity is needed. History has shown that the Fed needs a tool like Elastic Cash to inject liquidity into the financial system, since interest rates have proven too weak as a tool to inject large quantities of liquidity.  As we discussed in Section \ref{dollar}, this has led to the Fed  buying assets or propping up assets that were liable to crash in value. In doing so, the Fed is forced to pick and choose between market participants who get first access to the new liquidity that it provides. 

\item [Even partial implementations make sense] Central bankers should not be  attempting to directly reason about the demands for liquidity; they do not have enough data to make those decisions. But if they must make dramatic unilateral changes to the money supply, the scheme of Elastic Cash at the very least gives a fair way to do it by trading cashbonds along the lines I have suggested above. This can be done even if rule 1 is not enforced. The cashbond market would allow the Fed to inject money by buying cashbonds at auction. This removes the ability of the finance sector to control the flow of the new money.  It is also preferable to having the Fed buy treasuries or mortgage backed securities, because it disentangles the actions of the Fed from the needs of the Treasury and the mortgage industry. There is no need to tie increases in the money supply to increases in government spending or mortgage rates.

\item[Cashbond market $\neq$ US treasury market]  Cashbonds should not be confused with conventional government securities like US treasuries. These instruments are significantly different from each other, and one cannot make inferences about the cashbond market, which does not yet exist, based on the behavior of the US treasury market. Let me highlight some key differences:
\begin{enumerate} \item The issuance of cashbonds is controlled by strict and transparent rules, and is not tied to the spending of the US government. \item There is no analogue of debt ceilings, or any chance that the central bank will default. \item Cashbonds cannot be used as collateral for loans and cannot be transferred outside of the Elastic Cash market.
	\item Trade in cashbonds is not taxed.\end{enumerate}
For these reasons, I expect the cashbond market to be much more liquid than the current US treasury market, and I expect the risk-free rates of interest to be better behaved.

\item [Distribution on cashbonds] It is important for the functioning of Elastic Cash to maintain a large volume of outstanding cashbonds of varying durations. Ideally, we would like there to be broad participation in the cashbond market from all kinds of financial entities: banks, companies, pension funds, and individuals. Because these participants will be willing to trade at different durations, participation will be increased if a wide range of durations are available, and the market is liquid at all durations. Even though the cash authority only regulates the interest rate for duration $\tau$, this action will affect the rates for all durations. One would expect that banks and other sophisticated players will trade cashbonds of shorter duration, and perform the arbitrage necessary for information about demands for liquidity of all durations to  propagate to the shorter durations. I suspect that there is a principled way to choose the ideal distribution on durations of outstanding cashbonds, but I have not yet been able to convince myself about what it ought to be.

\end{description}

\section{Adopting Elastic Cash in the US Dollar system} \label{newdollar}
As discussed in Section \ref{dollar}, the Dollar system involves many different kinds of institutions that are currently creating instruments that can be exchanged for US Dollars. Changing the system is not going to be straightforward. However, I do believe that there is a path to making the change somewhat gradually, so that all the parties involved have time to  adapt to the new system. Here is a proposed sequence of steps for adopting Elastic Cash into the Dollar system:

\begin{enumerate}
\item The Fed begins to populate the cashbond market by gradually selling cashbonds of varying duration. Cashbonds are held at accounts maintained by the Fed, which allows the Fed to enforce that cashbonds cannot be transferred outside the cashbond market. At this point, cashbonds that expire are replaced according to the rules of Elastic Cash, but the risk-free rate of interest is allowed to float freely. I would expect this floating rate to converge close to the current Fed funds rate.
\item Once the market for cashbonds is running at significant scale, regulations should be enacted to curtail the private creation of US Dollars. This can be done gradually by raising the interest rate at which the Fed lends to private entities. At the same time, the Fed should begin to put upper and lower bounds on the risk-free rate by trading in the cashbond market. Once the cost of borrowing from the Fed far exceeds the risk-free rate in the cashbond market, private entities like banks will be incentivized to participate in the cashbond market and raise money there. The current creators of US Dollars can be handled as follows:
\begin{enumerate}
	\item  Commercial banks should be barred from creating new deposits that are not backed by cash reserves. Banks should fund new  lending activity by selling corporate bonds instead. As the interest rate for these bonds goes up, the risk-free rate of interest in the cashbond market will also go up via the trading in the cashbond market, and lead to the creation of new money that is funneled to loans. So, the money required to support loans will still be created via the new system. 
	\item  The Fed's repo facility and money market fund facility should be closed. \item The eurodollar market is, perhaps, a bigger problem, both because of its size and the fact that the institutions cannot be regulated by US law. Still, the Fed can wind down its swap lines with foreign central banks gradually, until eurodollars lose their Fed backing. Foreign central banks and governments should be allowed and  encouraged to participate in the cashbond market to obtain liquidity.
\end{enumerate}
\item The inevitable tantrums in the financial sector should be treated with stoicism. 
\end{enumerate}

It is an understatement that moving from our current system of private money creation to Elastic Cash would be a dramatic change. There are likely to be many challenges that need to be overcome to implement it, not least the resistance of the finance industry, whose raison d'{\^e}tre is to control the flow of money. Elastic Cash represents a significant loss of control for financial firms, and a democratization of the flow of money. For these reasons, it is perhaps more easily implemented in a cryptocurrency, as I discuss next.

\section{Elastic Cash in  cryptocurrencies}

A major advantage of Elastic Cash over conventional mechanisms for elasticity is that it can be implemented in a truly decentralized way, without \emph{any} trusted central authority. Bitcoin made a technological leap when it introduced the concept of a blockchain. Since then, a number of cryptocurrencies have emerged, with different ways to implement the blockchain. Any of these systems can be used to implement Elastic Cash, so here I will keep the discussion at a high level, only talking about how the blockchain can be utilized. Because Elastic Cash involves making significant changes to the money supply, I do believe that implementing it requires \emph{new} cryptocurrencies. I do not think it can be implemented using a layer built on top of Bitcoin, for example.

Here is how one can implement Elastic Cash on a blockchain at a high level:

\begin{enumerate}
\item At any point in time, each user of the cryptocurrency is known to hold some amount of money, as well as various cashbonds.
\item Users of the currency can announce transactions of money, as well as orders placed in the cashbond market. The orders can be placed with a specific expiry date.
\item Miners will add both money transactions and orders in the cashbond market to the next block of the blockchain. To implement the market in cashbonds:
\begin{enumerate}
\item Miners will act as market makers to map buy orders to sell orders and so execute the trade in cashbonds. There are some subtle issues that need to be addressed here. For example, a miner may be incentivized to choose some orders over others to include on the blockchain, insert their own orders in the final block, or choose to ignore some orders when acting as a market maker. In particular, miners can always insert orders to themselves obtain the spread between two crossing buy and sell orders, so it makes sense for miners to be paid this spread as a transaction fee to carry out their market making function. This removes the incentives to manipulate the orders that are added to the most recent block. 
\item Miners will also execute the algorithm to simulate the activities of the cash authority in the cashbond market. New money and cashbonds will be created according to the rules of the mechanism, and these will be traded with users based on the orders that have been added to the blockchain.
\end{enumerate}
\end{enumerate}

\section{Conclusions and Questions}
It is an exciting time to think about the technology of money. The US Dollar is experiencing a once-in-a-lifetime  contraction (see Figure \ref{m2}), and the demands for a stable global currency have never been larger. Elastic Cash is a broad scheme to enable elastic money. I have purposefully left the mechanism underspecified, because I believe that more work is required to understand the details and trade-offs involved in the particulars of the mechanism.

Here are some important questions that I feel remain unanswered:
\begin{enumerate}
	\item How should the market maker behave in the cashbond market? In the context of conventional currencies, can private entities function as market makers? In the context of cryptocurrencies, how should the algorithm be set up so that miners do not have an incentive to behave dishonestly when they are carrying out the role of market maker? 
	\item What style of auction would give the best results for the cashbond market? 
	\item What is the ideal target distribution on cashbonds? If the cashbonds are concentrated on very short durations, this gives the most power for the mechanism to inject large quantities of money, but it also means that the market loses information about the demand for liquidity over long durations. So, there is a trade-off between various choices for distributions on durations.
	\item How can we gradually transition the current US Dollar system to such a mechanism? The steps I discussed in Section \ref{newdollar} are likely to be difficult to execute. Perhaps there is a way to use cashbonds and incentivize the large players in the financial system to adopt Elastic Cash without being forced to do it. What is needed is a mechanism to transition to the new mechanism!  
	\item How should we expect the free floating rate curve $\mathsf{rate}(t)$ to behave as a function of $t$ during normal times? I would expect this function to be monotone, but I am not sure how to reason about it beyond that.
\end{enumerate}

\section{Acknowledgements}
Thanks to Paul Beame, Siddharth Iyer, Travis Kriplean, James Lee, Noam Nisan, Darcy Rao, Tim Roughgarden, Eli Ben-Sasson, Oscar Sprumont, Michael Whitmeyer and Amir Yehudayoff for many helpful and entertaining conversations about money. 

\label{newcrypto}

%

\end{document}